\documentstyle[aps,eqsecnum,twocolumn,epsf]{revtex}
\begin{document}

\title{\bf Kinetics of shape equilibration for two dimensional islands}

\author{Pablo Jensen$^*$ (a), Nicolas Combe (a), Hern\'an Larralde (b), 
Jean Louis Barrat (a), Chaouqi Misbah (c) and Alberto Pimpinelli (d)}
\address{(a) D\'epartement de Physique des
Mat\'eriaux, UMR CNRS 5586, Universit\'e Claude Bernard Lyon-1, 69622
Villeurbanne C\'edex, FRANCE;\\ (b) Instituto de F\'{\i}sica, Lab. de
Cuernavaca, Apdo. Postal 48-3, C.P. 62251, Cuernavaca, Morelos, MEXICO
\\ (c) Laboratoire de Spectrom\'etrie Physique, Universit\'e Joseph
Fourier (CNRS), Grenoble-1, B.P. 87, 38402 Saint-Martin d'H\`eres
C\'edex, FRANCE \\ (d) LASMEA, Universit\'e Blaise Pascal Clermont-2, Les
C\'ezeaux, 63177 Aubi\`ere C\'edex, FRANCE}

\maketitle
\narrowtext

\begin{abstract} 
We study the relaxation to equilibrium of two dimensional islands
containing up to 20000 atoms by Kinetic Monte Carlo simulations.  We
find that the commonly assumed relaxation mechanism - curvature-driven
relaxation via atom diffusion - cannot explain the results obtained at
low temperatures, where the island edges consist in large
facets. Specifically, our simulations show that the exponent characterizing 
the dependence of the equilibration time on the island size is different
at high and low temperatures, in contradiction with the above cited assumptions.
Instead, we propose that - at low temperatures - the relaxation is limited 
by the nucleation of new atomic rows on the large facets : this allows us to
explain both the activation energy and the island size dependence of
the equilibration time.  
\end{abstract}


\section{Introduction}
There is a continued interest in the understanding, description and
control of structures at the nanometer scales
\cite{generale,www_nano,fluelli,duxbury,metiu,averback,LECD}.  This is
partially due to technological applications of nanostructures, and
partially to the fundamental interest of understanding how macroscopic
concepts can (or cannot) be extrapolated down to these scales. On the
technological side, a controlled preparation and conservation of these
structures demands a comprehension of their time evolution, which
could be rapid due to the small scales involved (typical structures
contain some hundred atoms).  From the fundamental point of view, it
is interesting to investigate how the theoretical tools which have
been developed to deal with the kinetic evolution of macroscopic
objects (size larger than a micrometer) by Herring, Mullins and
Nichols \cite{nichols}, which are based on coarse-grained, continuous
equation, can be used at the nanometer scale.  For example, one could
wonder whether sintering of ceramic or metallic nanopowders can be
analyzed with these classic tools since it is not clear that
macroscopic concepts such as curvature, chemical potential,
etc. should retain their relevance when dealing with structures
containing only few atoms.

Here, we focus in the {\it dynamics} of equilibration of
{\it two dimensional} (2D) nanocrystallites starting in an out of 
equilibrium shape.  Mullins'
theory of shape relaxation is based on the curvature being well
defined. Then, two questions arise. Can one use partial differential
equations to study the shape relaxation of faceted nanocrystals? Can
one use them at all far from the thermodynamic limit---i.e. for small
crystallites?  Studying the validity of the partial differential
equations approach at various length scales and temperatures is
important since this formalism is also used by experimentalists to
derive diffusion constants \cite{drechsler,bonzel} or interpret their
data \cite{jeffers}. Some workers \cite{jullien,eggers} have also
used this approach as a black box to calculate the time evolution of
different structures. Our
study is related to one of the major problems of equilibrium surface
physics : the comprehension of the relaxation of a perturbed surface
profile, below its roughening temperature $T_R$
\cite{vipi}. Above $T_R$, the surface (of an infinite volume
crystal whose ratio surface area/volume is finite) is rough. This is
equivalent to say that the step free energy vanishes identically
(steps can be created at no free energy cost), and it also implies
that the surface free energy is an analytic function of the local
slope. Below $T_R$ the surface is smooth. This means that the step
free energy is non-vanishing, and that the surface free energy is
non-analytic. Indeed, below $T_R$ the surface free energy displays
cusps at particular orientations, which coincide with facets in the
equilibrium shape. In the thermodynamic limit, the chemical potential
of a crystal is defined from the Gibbs-Thomson relation, $\mu = \Omega
K $, in terms of the surface crystal curvature $K$. On facets, the
curvature is ill-defined, and the chemical potential is fixed by the
curved part surrounding the facet. For the relaxation of a 2D surface 
(the surface of a bulk crystal), it is generally believed that Mullins'
treatment is correct above $T_R$. Below this temperature, different 
approaches have been proposed \cite{villa}.

Molecular dynamics (MD) simulations of the coalescence of
three-dimensional clusters containing roughly 1000 atoms, have shown
that the relaxation kinetics is slower than predicted by Mullins'
theory \cite{llewis}. However, MD simulations are still limited in
computation time (no more than $\sim 10 ns$) and it is therefore
difficult to follow the coalescence at temperatures not too close to
the melting temperature (see also \cite{duxbury,averback}). An
alternative method consists in using Kinetic Monte Carlo (KMC)
simulations which allow an incomparable larger range of time
studies. As a first attempt, we have chosen to study a two-dimensional
(2D) system, namely the relaxation kinetics of 2D faceted islands
supported on a triangular lattice.

The basic idea is the following : we start with a island with a shape
clearly not an equilibrium one (e.g. with an $x$ side 10 times longer than
the $y$ side) and anneal it at a given temperature. Indeed, the
perimeter free energy dictates the equilibrium shape, which is the one
that minimizes the island free energy at given volume (here we expect
hexagonal shapes because of the lattice geometry).  We assume that the
island has relaxed when the aspect ratio (defined as the ratio of the
$x$ and $y$ gyration radii, $\alpha = \rho_x/\rho_y$) becomes 1. We
then monitor the {\it kinetics} of the relaxation process, which
depends on the precise pathway chosen by the island to change its
shape. We are particularly interested in the influence of {\it facets}
in the relaxation kinetics.  For this reason, we perform simulations
at several temperatures: at high temperatures, where the islands
contour is clearly rough and only small facets can be distinguished;
at low temperatures, where clear-cut and persistent facets are
apparent. This point is quite delicate, because a 2D object---whose
contour is a line, not a surface---is not expected to show facets at
any temperature {\it in the thermodynamic limit}. Indeed, a line is
rough at all non-zero temperatures, and the line tension $\gamma$ 
($J m^{-1}$) is analytic
for all orientations \cite{vipi}. However, facets do appear at $T=0$,
and at low $T$ one would expect the persistence length of a facet (the
average distance between kinks) to be quite large. If it is larger
than the island side, then facets are indeed expected, as we
observe. In other words, creating a kink costs a finite energy, which
is always compensated by the entropy gain when the length of the line
goes to infinity (cf. Landau's argument for the non-existence of phase
transitions in 1D). As long as the line is finite, facets occur. One
could then guess that they affect the kinetics, even for 
$T > T_R = 0$.

Our main conclusion is that equilibration of an island's shape is a
non-universal process, in which the time evolution of the shape does
not obey scaling, while it strongly depends on temperature, and thus
on system-dependent features like the energy scale $E$. Scaling
relations can be found for the relaxation (or equilibration) time, as
a function of temperature and island size.  Indeed, two regimes with
two different scaling forms are born out by the simulations, at high
and low $T$, respectively. We tentatively attribute these two regimes
to the absence and presence of facets, respectively.
 
\section{Monte Carlo and Partial Differential Equation approach}

\subsection{Kinetic Monte Carlo simulations}
\label{kmc}

\begin{figure}
\centerline{
\hbox{
\epsfxsize=5cm
\epsfbox{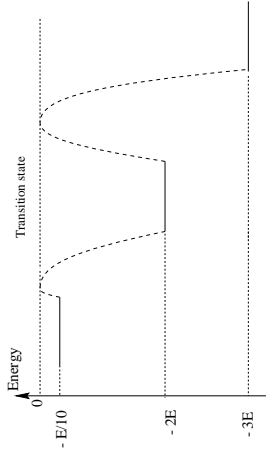}}
}
\caption{Potential energy of an atom diffusing along the island edge
when it has 1 (energy : -E/10), 2 (energy : -2E) or 3 (energy : -3E) neighbors.
The atom energy is supposed to depend only on the number of first neighbors
and the transition state is assumed to lie at the same energy for all 
jumps, taken here as the origin of energies. As a consequence, the energy
barriers for diffusion are equal to the potential energies of the atoms,
ensuring detailed balance (see text).}
\label{energy_profile}
\end{figure}

We perform ``standard" kinetic Monte Carlo (KMC) simulations on a
triangular lattice. We assume that the potential energy of an atom is 
proportional to its number of neighbors, and that the {\it kinetic barrier}
for diffusion is also proportional to the number of {\it initial} neighbors, 
regardless of the {\it final} number of neighbors, i.e. after the jump (see
Fig. \ref{energy_profile}). This is of course a huge simplification, 
which is however aimed here
at describing the global evolution of a model island. In other words,
we do not wish to study any particular system but rather to
investigate properties which should not depend on the details of
atom-atom interaction. Therefore, we use a simple kinetic model
containing as few parameters as possible (only one, the ratio $E/k_B
T$ where $E$ sets the energy scale ($E=0.1$ eV throughout the paper),
$k_B$ is the Boltzmann constant and $T$ the absolute
temperature). Comparing with recent ab-initio calculations
\cite{bogicevic} for the Al(111) surface, we note that our one-barrier
assumption does give the good order of magnitude of the relative jump
frequencies for the different hopping processes of interest here. We
also exclude any explicit ``Ehrlich-Schwoebel" barrier \cite{es} for
atoms hopping around corners, although the occurrence of atoms with a
single neighbor is treated in a special way (see below). The kinetic
barriers for some jumps are shown in Fig. \ref{jumps}. In the same
spirit as ours, a similar but slightly more complicated model has been
used recently by Metiu's group with the scope of obtaining
system-independent information on island {\it diffusion} on a surface:
these authors investigate the existence of a ``universal" size
dependence of the island diffusion constant. They conclude that such
universality is not observed, and we observe a similar phenomenon for
island {\it equilibration}.

The time evolution of the island shape is obtained by the following
algorithm. We first calculate the following quantities :
$\delta_1=\exp[-E/(10k_BT)]$, $\delta_2=\exp[-2E/(k_BT)]$ and
$\delta_3=\exp[-3E/(k_BT)]$ which represent the relative weights for
the jump probabilities for atoms with respectively 1, 2 or 3 neighbors
(atoms with more neighbors simply do not move : see an explanation of the 
precise forms of the different $\delta_i$ below). Then, in each
iteration, we calculate the probability to move an atom with $i$
neighbors as : 

\begin{equation} 
p_i={(6-i)n_i\delta_i \over \sum_{i=1}^{3}{(6-i)n_i\delta_i}} 
\end{equation} 

where $n_i$ is the
total number of atoms having $i$ neighbors. We choose randomly one of
the atoms with the appropriate number of neighbors and move it in a
random direction. The time is increased at each iteration by

\begin{equation} 
dt=\left({\nu_0}\sum_{i=1}^{3}{(6-i)n_i\delta_i}\right)^{-1}
\end{equation} 

where $\nu_0$ is a Debye frequency (we have taken $\nu_0 = 10^{13} s^{-1}$). 
To check that the law of detailed balance is satisfied, one can refer to
Fig. \ref{energy_profile} : the probability for an atom to jump from 
a site having $n$ neighbors to a site having $p$ neighbors is $\delta_n$
while the opposite transition has a probability $\delta_p$. Their ratio is
$\delta_n$/$\delta_n = exp(-E/(k_BT)(n-p))$, i.e. equal to the energy
difference of the initial and final configurations, as required by the
law of detailed balance (the particular case when $n$ or $p$ is equal
to 1 can be analyzed in the same way). This algorithm is very
fast \cite{kalos,kang,accemetiu} since all iterations contribute to
the evolution (there are no rejected moves). One peculiarity of this
model is the treatment of atoms having one single neighbor: $\delta_1$
is much larger than what one could expect from the general rule
$\delta_n=\exp[-nE/(k_BT)]$: indeed, we let
$\delta_1=\exp[-E/(10k_BT)]$ instead of
$\delta_1=\exp[-E/(k_BT)]$. This is to ensure that singly-bonded
atoms, which are in some sense in a ``transition state", rapidly go
into some physically reasonable position, i.e. one having 2 or more
neighbors.  Note also that detachment of atoms from the islands is
forbidden here: equilibration is only due to mass transport along the
island contour. This is clearly different from Ostwald ripening where
islands evolve in equilibrium with a two-dimensional adatom gas. We note that
a recent experimental study by Stoldt et al. \cite{stoldt} has shown that 
supported Ag two dimensional
islands do indeed relax via atomic diffusion on the island edge, without
significant contribution from exchange with a two-dimensional adatom gas. A last
remark on the algorithm used here : we do not allow atoms having more than
3 neighbors to move. In some sense, they have an infinite potential energy.
Since our potential energy is not very realistic anyway, this hypothesis 
allows to simplify and accelerate the simulations. The key point is that
our results are particularly interesting at {\it low} temperatures, where
including the possibility for atoms with 4 neighbors to jump would make
no significant difference in the kinetic evolution of the island because
at these temperatures their jumping is vanishingly small.

\subsection{Partial Differential Equation approach}
\label{calcul}

A complementary approach at predicting the evolution of a crystal
shape at a temperature higher than the roughening transition, consists
in coarse-graining the crystal profile, in order to treat it as a
smooth function $h(x,t)$, and in writing down its time evolution in
the form of a partial differential equation, whose form depends on the
physical situation of interest.  The situation when matter transport
is assured by adatom diffusion along the surface, has been originally
considered by Mullins and coworkers \cite{nichols}, for studying small
deformations of an infinite planar surface. The case of a finite,
closed "surface" --- the island contour --- is somewhat more subtle, and
we give the derivation in some detail. The evolution equation has in
general then the form of an equation for $s$ the curvilinear
coordinate or arclength. On purely geometrical grounds, it can be
 shown \cite{miva} that this equation can be written as an
evolution equation for the curvature $K(s,t)$, of the form

\begin{equation} 
{\partial K \over \partial t }= -\left[{\partial^2
\over \partial s^2 } +K^2 \right] v_n - {\partial K \over \partial s
}\int {\rm d}s' Kv_n 
\label{CCE} 
\end{equation} 

where $v_n$ is the normal velocity of the interface. The latter is
fixed by the physics of the problem. In our case, when edge
diffusion is the relevant physical process determining the relaxation
of the shape, the equation for $v_n$ must have the form of a
conservation equation for the island area 

\begin{equation} 
{v_n }= -{\partial{\bf j} \over \partial s } 
\label{CE} 
\end{equation} 

The edge diffusion current ${\bf j}$ is given by the gradient of
the local excess chemical potential as in Fick's law :

\begin{equation}
{\bf j}=-{\tilde D \over k_B T} {\partial \Delta\mu\over \partial s} 
\label{FL}
\end{equation} 

where $\tilde D$ is a (collective) perimeter diffusion coefficient.

The excess chemical potential is in turn related to the local
curvature $K(s)$ through the Gibbs-Thomson relation: 

\begin{equation}
\Delta\mu = - { \gamma a^2} K 
\label{GT} 
\end{equation} 

where $\gamma$ is the line tension (that for simplicity is assumed
isotropic here), and $a$ the depth of the outer layer within which
mass transport takes place (one lattice spacing in our
case). Eqs. (\ref{FL}) and (\ref{GT}) yield 

\begin{equation} 
{\bf j} =- {\tilde D \gamma a \over k_BT} {\partial K \over \partial s }
\label{EM1}
\end{equation} 

and the corresponding evolution equation (\ref{CE}) is given, in our
two-dimensional situation, by:

\begin{equation} 
v_n = {\tilde D\gamma a^4 \over k_BT} {\partial^2 K
\over \partial s^2}  
\label{EM2} 
\end{equation} 

Equations (\ref{CCE}) and (\ref{EM2}) must be solved simultaneously
for the island shape: the results will be shown in section
\ref{path}. The equilibrium solution $v_n=0$ obviously is a
constant-curvature shape, that is a circle. Also, note that equations
(\ref{CCE}) and (\ref{EM2}) are invariant by a rescaling
$s\rightarrow\lambda s$, $t\rightarrow\lambda^4 t$ \cite{nichols}, so
that the equilibration time of a deformed island of size $L$ is
expected to be proportional to $L^4$ or $N^2$ where $N$ is the number
of atoms in the island.

\section{Monte Carlo simulation results}

To study the influence of the facets on the coalescence kinetics, it
is interesting to study the time evolution of the aspect ratio and
the size, as well as the temperature dependence of the equilibration time
$t_{eq}$. We recall that $t_{eq}$ is defined as the time needed for
the island to reach its equilibrium shape. In practice, we take
$t_{eq}$ as the first time when the aspect ratio $\alpha$ defined
above becomes less than 1. Each point is the average of several runs
(up to 200 for the smallest islands).

\subsection{Island morphology}
\label{morpho}

Fig. \ref{perimt}a shows the time evolution of the perimeter of a 6250
atoms island at 500 K ($E/k_B T = 2.3$). It is clear that the shape
evolution occurs with rough island borders.

Fig. \ref{perimt}b shows the time evolution of the perimeter of an
island containing 6250 atoms at 83 K ($E/k_B T = 14$). At this
temperature facets are apparent throughout the evolution.

A more precise comparison of the presence of facets at the two
temperatures studied above is given in Fig. \ref{6250fac}. It is
apparent that facets are present at 83 K, in contrast to the high
rugosity observed at 500K.

\subsection{Dependence of the equilibration time on island size}
\label{size}

The continuous analysis of section \ref{calcul} predicts $t_{eq}\sim
N^2$ as a function of the number of atoms $N$ inside the island, for
any temperature. Indeed, the
numerical solution of the full, non-linear equations (\ref{CCE}) and
(\ref{EM2}) appears to agree with this prediction. Fig.  \ref{teqfN}a
shows the size dependence of $t_{eq}$ for different temperatures as
given by the simulations. The simulation results agree with
$t_{eq}\sim N^2$ {\it only at high temperatures}.  Below 250 K, it is
clear that $t_{eq}$ increases slower than $N^2$, and the lower the
temperature, the smaller the exponent. One can also notice
(Fig. \ref{teqfN}b) that the local exponent for low temperatures
approaches 1 for the highest island sizes. This is analyzed in
Sec. \ref{scaling} where we give an attempt at deriving a scaling
relation describing the two regimes. It should also be noted that
extrapolating the different curves for very high values of the island
size leads to an apparently absurd conclusion : very large islands
do equilibrate faster at lower temperatures. This is a immediate
consequence of the higher size exponents found for the highest
temperatures. To avoid a paradox, we must admit that there exist a
crossover from high to low temperature behavior for a given size
that depends on the temperature. Therefore, even at 83 K (highest
curve), for large enough islands, one should recover the $t_{eq}\sim N^2$
regime. The scaling analysis presented below explains this crossover.

\subsection{Dependence of the equilibration time on temperature}

Fig. \ref{teqT}a shows that $t_{eq}$ rapidly increases as temperature
decreases, in roughly the same way for all the island sizes. The
equilibration time is not exactly a thermally activated quantity,
since there is a clear curvature in its Arrhenius plots, as shown in
Fig. \ref{teqT}b: the local activation energies increase from roughly
0.3 eV at high $T$ to 0.4 eV at low temperature. This represents
respectively 3 and 4 times the energy needed to break a single bond. A
tentative interpretation of these values is given below in Sec.
\ref{scaling}.

\subsection{Precise kinetics of the relaxation}
\label{path}

We have seen that the size dependence of the equilibration time obeys
the scaling predicted by the linearized equations only at high
temperatures. It is interesting to check whether the full solution of
equations (\ref{CCE}) and (\ref{EM2}) agrees with the high-temperature
behavior of the MC simulations, when no facets are apparent, and the
island looks rough and rather isotropic. Fig. \ref{kinetics}, where
the aspect ratio is plotted as a function of the reduced time
$t/t_{eq}$, shows that no agreement is found, at any of the studied
temperatures. Indeed, this is {\it a posteriori} not surprising, since
the MC results do not seem to obey any scaling relation, or maybe only
at high temperature, and it is then obvious that the ``universal''
description given by continuous equations does not apply. It is
nevertheless a little surprising that the continuous description,
which agrees with simulations in the case of a planar surface above
the roughening temperature, does not seem to set a limiting behavior
valid for very large sizes ($N\to\infty$), nor to provide the scaling
form which seems to appear in the simulation results at high
temperatures.

We can think of (at least) three explanations of this
observation. First, it could be argued that this is an effect of the
edge tension.  Indeed, in writing the constitutive equation
(\ref{EM2}) we have assumed that $\gamma$ is isotropic. This is
clearly not the case in the simulations since we take a triangular
lattice, and the energy of the facets depends on the
orientation. We are currently performing a numerical integration of
equations (\ref{CCE}) and (\ref{EM2}) including anisotropic line tension
to further investigate this point.

Second, one could question the adequacy of the continuous treatment to
describe the detailed path to equilibrium, even at high temperatures,
for small clusters. Indeed, one could argue that the macroscopic
concepts of curvature, chemical potential, etc. are not adapted to
deal with {\it nanometric} objects containing few (less than 10000)
particles.

Finally, it is possible that the simulations are not adapted to agree
with the continuous theory, in the sense that our interatomic potential
is too crude to give a reasonable kinetic path to equilibrium. It is
clear that the assumption that the transition probability depends only
on the {\it initial} state is not generally correct.  As has been
argued above (section \ref{kmc}), one would expect such a rough 
potential to reproduce a
{\it universal} exponent (as is observed for $t_{eq} \propto N^2$ at
high temperatures) but not necessarily a detailed time evolution, if
the latter is non-universal.

\section{A Scaling Argument}
\label{scaling}

Our MC results show that two different regimes - at low and high
temperatures - can be identified. At first sight, this is surprising 
since
the only occurrence that could separate high from low temperatures is the
roughening transition, which, strictly speaking, takes place at $T=$0 K 
in 2 dimensions --- that is, for a one-dimensional "surface". However, facets do
not disappear suddenly as the temperature is raised from
$T=0$. Indeed, a ``persistence length'' of facets can be defined as the
equilibrium value of the distance between {\it kinks} along the step edge.
At equilibrium and at low temperatures, we can consider an ``ideal gas'' of
kinks, whose density (number of kinks per unit length) is given by the
formula \cite{bcf}
\begin{equation}
n_{\rm kink}=2\exp(-\beta W)/a
\label{kinkden}
\end{equation}
where $W$ is the kink creation energy and $a$ the lattice spacing. Indeed,
to form kinks one has to take an atom out of the step edge, and to place 
it
anywhere along the step. On a triangular lattice, in doing so one looses 4
nearest-neighbours bonds, and gains 2. A net balance of 2 broken bonds
results. In the process, 4 kinks have been created (each atom counts for 2
kinks), so that the energy cost per kink is
\begin{equation}
W=E/2\;.
\end{equation}
The factor of 2 in (\ref{kinkden}) comes from the fact that kinks always
appear and disappear in pairs---in other words two types of kinks, 
positive
and negative exist, of equal number. Then, the equilibrium distance 
between
kinks is 
\begin{equation}
\ell_0=a\exp(\beta E/2)/2,
\end{equation}

which diverges as $T$ goes to 0. At a given temperature, a step looks
straight (free from kinks) over lengths of the order of $\ell_0$. An 
island is thus bound to look faceted as long as $\ell_0$ is larger than the 
island linear size $L$. Then, the approximate equality

\begin{equation}
\exp(E/(2 k_B T_c)/2\approx L_c/a
\end{equation}

gives the ``crossover size'' $L_c$ (at fixed temperature) or the 
``crossover temperature'' $T_c$ (at fixed size) for the crossover between
the high (rough) and low (faceted) temperature regimes. A comparison with
Fig.(5b), where the high temperature regime corresponds to $t_{eq}\sim 
N^2$, while the low temperature regime corresponds to $t_{ eq}\sim N$, shows 
that this criterion is not too bad: the formula predicts a crossover 
temperature for an island of size $N=500$ ($L\approx 22$) of approximately 240 K, in 
good agreement with the simulation data. Note that for smaller sizes than about
$N\approx 100$ the simulations do not show a well defined low temperature
behaviour. We attribute this to the large importance of geometric kinks
(imposed by the fact that the step closes on itself) over thermal kinks 
for these small sizes. Indeed, we claim that the low temperature regime is 
ruled by the equilibration of the spatial distribution of thermal kinks: the
initial shape creates a strongly inhomogeneous distribution of these 
kinks, which then diffuse to achieve spatial uniformity, and thus equilibrium.
Kinks diffuse by emitting atoms, so that we conclude that atom emission 
from kinks is the limiting  kinetic step determining the low temperature
behaviour of $t_{eq}$. 

Based on this assumption, we can give a scaling argument that
reproduces well the observed $t_{eq}$ as a function of $N$ and
$T$. The argument is similar to that used by Bales and
Zangwill \cite{bz} and Pimpinelli et al.  \cite{pimpicine} to discuss
step roughening and smoothening during growth and at
equilibrium. Indeed, it amounts to performing a linear stability
analysis, and computing the relaxation time of a perturbation of given
amplitude and wavelength. For sake of clarity, we start from the 
discussion of the high temperature behaviour, which is well known from Mullins work.

Physically, we assume that there are two
ingredients determining the relaxation : the thermodynamical ``force"
which drives the relaxation (here, the excess curvature) and second, 
the kinetic factors which determine the
{\it rate} of the equilibration. At all temperatures,
curvature effects are relevant, but we assume that the kinetics change
due to the presence of facets (or, equivalently, to the low
concentration of kinks). The transition takes place, as stated above,
when $\ell_0 \approx L_c$. Here is the mathematical translation of this
idea.

\subsection{High temperatures}

Let $\delta_q$ be the amplitude of a perturbation of wave vector $q$ of the
island perimeter with respect to the equilibrium shape. The
curvature effect (Gibbs-Thomson) opposes the increase of the
deformation. The rate of decrease depends on the appropriate kinetic
process which limits transport of matter from high to low chemical
potential regions. Let $n_{eq}$ be the equilibrium atom density along
a reference island edge with the equilibrium shape. Then, at high
temperature, a deformation of local curvature $K$ results in an excess
chemical potential $\Delta\mu\sim\gamma q ^2\delta_q$ as in equation
(\ref{GT}). In turn, this creates an excess atom density
$n_{exc}=n_{eq}\exp[\Delta\mu/(k_BT)]\approx
n_{eq}[1+\Gamma q^2\delta_q]$, where $\Gamma=\gamma/(k_BT)$.  Then, edge
atoms flow away from the deformation, whose amplitude decreases at a rate
proportional to the divergence of the mass current :
\begin{equation}
\dot\delta_q \approx - {1\over\tau^*} \nabla^2 (n_{exc}-n_{eq})\approx -
{n_{eq}\over\tau^*}q^2 \times {\Gamma q^2}  \delta_q
\label{nexcess}
\end{equation}
where $\tau^*$ is the typical timescale of the appropriate kinetic
process which is responsible of matter transport.

A more detailed justification of this expression can be found in Mullins
\cite{}, Bales and Zangwill \cite{bz} and Pimpinelli et al. 
\cite{pimpicine}.

Defining the equilibration time $t_{eq}$ by writing
$\dot\delta_{q=1/L}\approx -\delta_{q=1/L}/t_{eq}$ gives
\begin{equation}
t_{eq}\approx L^4{\tau^*\over n_{eq}\Gamma}\approx N^2{\tau^*\over
n_{eq}\Gamma} \;.
\label{teqN}
\end{equation} 
Mullins equation is recovered if one assumes that atom edge diffusion 
limits
the kinetics, so that 
\begin{equation}
{1\over\tau^*}\approx {D}\;.
\label{HT}
\end{equation}
The atom equilibrium density can be obtained from the detailed balance
at the kinks: $Dn_{ eq}=\nu_{ kink}$, where $\nu_{
kink}=\nu_0\exp{[-3E/(k_bT)]}$ is the rate of atom emission from kinks
and $Dn_{\rm eq}$ is the atom flux to the kinks
\cite{bales,prbevap}. Thus,
\begin{equation}
n_{ eq}=\nu_0/D_0\exp{[-E/(k_bT)]}\;.
\label{neq}
\end{equation}

Inserting (\ref{HT}) and (\ref{neq}) in Eq. \ref{teqN} yields, in the
limit of high temperatures,

\begin{equation}
t_{eq} \approx {1 \over \Gamma \nu_0} N^2 \exp{[3 E/(k_bT)]}
\label{teqHT}
\end{equation}

This prediction reproduces the $t_{eq}\sim N^2$ scaling of the
continuum theory, and it is in very good agreement with the simulation
results obtained at high temperatures both for the temperature
dependence and for the size dependence
(Fig. \ref{teqfN}. Indeed, at high temperature the equilibration
time shows an activation energy of approximately
$3E$ (Figs.  \ref{teqT}), and $t_{eq}$ behaves $\sim N^2$ in this regime.

\subsection{Low temperatures}

The low temperature regime sets in, for a given crystal size, when the
equilibrium distance between kinks becomes of the order of the linear 
size of
the crystal, and straight step portions appear. The (thermal) kink density
then becomes a relevant concept. When the crystal is deformed from the
equilibrium shape, the kink density is increased where the facets are
shrunk, and decreased where they are streched. On removing the 
constraint,
the kink density tends to equilibrium and seeks spatial uniformity. If the
equilibrium facet size is $L\approx N^{1/2}$, and a shape 
deformation of order
$\delta \ell$ is introduced, the kink density unbalance is approximately
$\delta \ell/L^2$. Then, the perturbation relaxes as
\begin{equation}
\dot{(\delta\ell)} \approx - {1\over\tau^{**}} 
 \times {1 \over \L^2}  \delta \ell \approx  - {1\over\tau^{**}} 
 {1 \over N}  \delta \ell.
\label{nexcess2}
\end{equation}
The relaxation proceeds by moving  a whole row of atoms from
a short to a long facet; diffusion is fast on facets, and the process is
limited by nucleation of the new row, that is, by the rate of atom
encounters $Dn_{eq}^2$. Then,
\begin{equation}
{1\over\tau^{**}}\approx Dn_{eq}^2=\nu_0^2/D_0\exp[-4E/(k_BT)]
\label{LT}
\end{equation}

Inserting (\ref{LT}) and (\ref{neq}) in Eq. \ref{nexcess2} yields, at low
temperatures,

\begin{equation}
t_{eq} \approx {D_0 \over  {\nu_0}^2} N \exp{[4  E/(k_BT)]}
\label{teqLT}
\end{equation}

Again, the activation energy predicted here is in good agreement with
the low temperature limit observed in the simulations
(Fig. \ref{teqfN}). The scaling $t_{eq}\sim N$ is less clearly seen in
the simulations (Figs.  \ref{teqT}). However, the simulations show
that the lower the temperature, the lower the size exponent, and if
$N$ is not too small, $t_{eq}\sim N$ is consistent with our
results. When $N$ is smaller than about 100, $t_{eq}$ seems to
increase faster than linearly. At such small sizes, facets are always very
short, and it is likely that an intermediate behaviour between mass
transport and facet nucleation rules the relaxation.

\subsection{Discussion}

The scaling argument we propose nicely reproduces the  results of our
simulations and leads to a reasonable physical picture of the 
equilibration, consistent with the observed morphologies and kinetics 
(presence of facets, rapid completion of atomic rows \ldots). 

Even more, our results can be used to estimate the behaviour of the 
diffusion coefficient of a cluster as a function of the cluster size 
$N\approx L^2$, by means of another scaling argument. In order to diffuse 
over a length $\ell$, a number of atoms of order $\ell L$ have to be
transferred from one side of the island to the opposite side. The time 
needed to do this is of the order of the time $t_{eq}(L)$ needed to 
equilibrate a fluctuation of linear size $L$ and mean square amplitude 
$\ell^2\approx L/(\beta\gamma)$ \cite{vipi}. Therefore, from the knowledge 
of $t_{eq}(L)$ we can know the diffusion coefficient $D(N)$ from the 
Einstein relation

\begin{equation}
D(N)t_{eq}(N)\approx \ell^2 \sim N^{1/2}.
\end{equation}

If we assume that our high-temperature result $t_{eq}(N)\sim N^2$ holds, 
we find

\begin{equation}
D(N) \sim N^{-3/2}.
\label{threehalves}
\end{equation}

If we assume that our low-temperature result $t_{eq}(N)\sim N$ holds, we 
find

\begin{equation}
D(N) \sim N^{-1/2}.
\label{onehalf}
\end{equation}

Equations (\ref{threehalves}) and (\ref{onehalf}) can be compared with 
the results 
of the simulations of island diffusion of Metiu and coworkers': on a 
(001)-type substrate they find that the size-dependent diffusion constant
$D(N)$ of 2D islands varies as $D(N)\sim 1/N^{1.52}$ at high temperature,
and as $D(N)\sim1/N^{0.62}$ at low temperature \cite{metiumrs}. Of course, 
different equilibration processes would lead to different
$t_{eq}(N)$s and then to different behaviours for $D(N)$. This might explain
the different results obtained by Bogicevic et al. \cite{metiudiff}
for islands diffusing on a (111) substrate but with energy
barriers for the jumps different from those assumed here.

\section{Summary, Perspectives}

The relaxation to equilibrium of 2D islands containing up to 20000
atoms shows unexpected features.  Our results show that there is no
"universal" size exponent for island equilibration, a result similar
to that found by Metiu's group for island {\it diffusion}
\cite{metiudiff}. We are now studying the case of 3D clusters to 
check both the scaling
of the equilibration time with the size of the particle and the
precise kinetic path followed to reach equilibrium.  This is done by
KMC simulations and an analytical approach.

\vspace{1cm}

$^*$ e-mail address : jensen@dpm.univ-lyon1.fr

\newpage

\begin{figure}
\centerline{
\hbox{(a)
\epsfxsize=5cm
\epsfbox{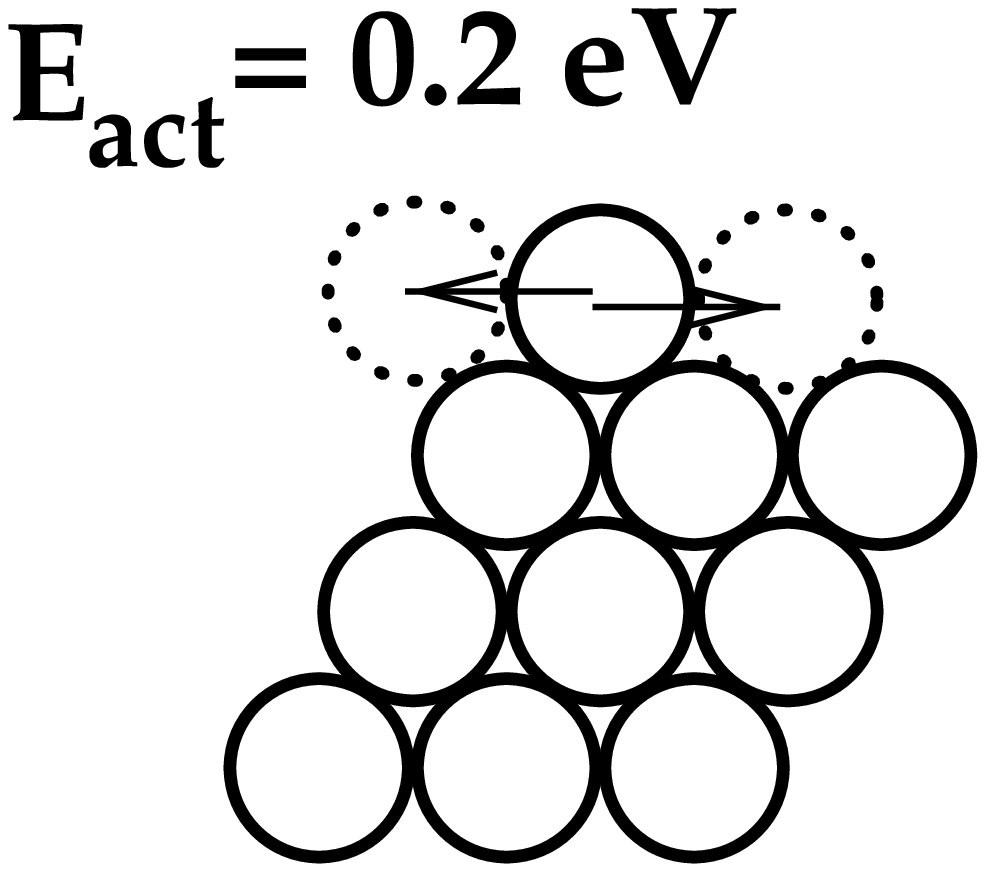}
}}
\end{figure}

\begin{figure}
\centerline{
\hbox{(b)
\epsfxsize=5cm
\epsfbox{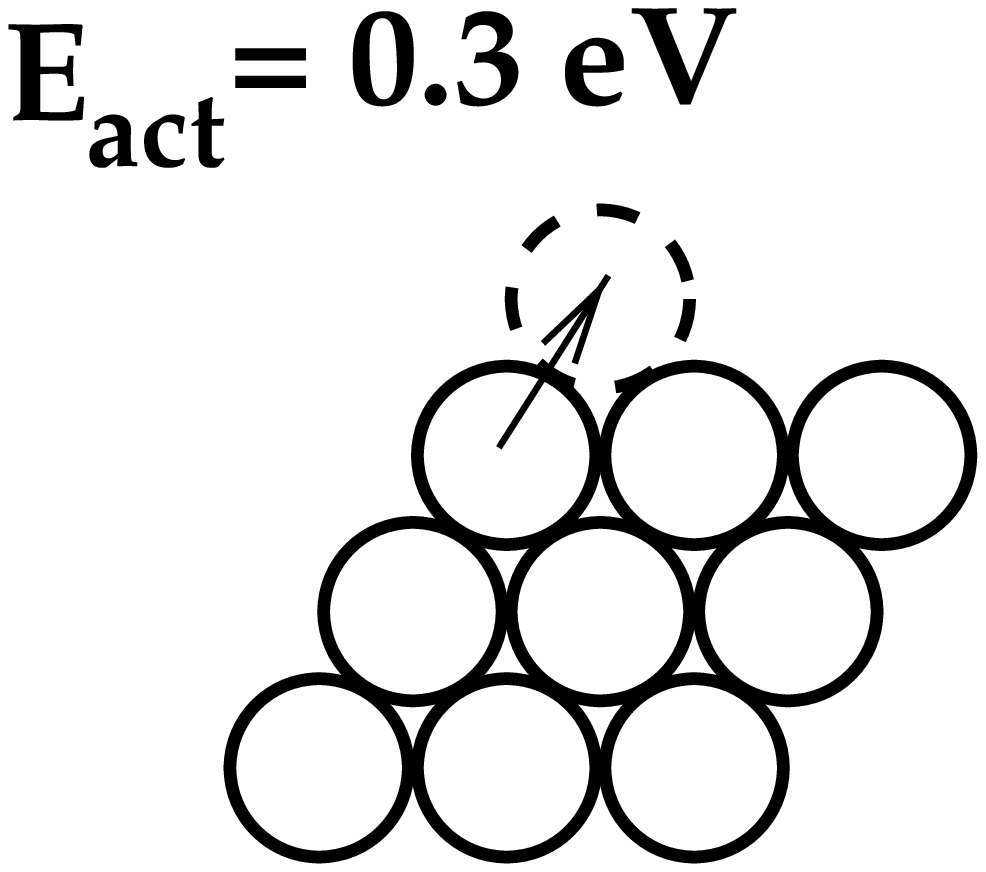}}
}
\caption{Examples of activation energies for different atomic jumps on
the island edge. Note that detachment of atoms from the island is
explicitly forbidden in the simulations.}
\label{jumps} 
\end{figure}

\begin{figure}
\centerline{
\hbox{(a)
\epsfxsize=5cm
\epsfbox{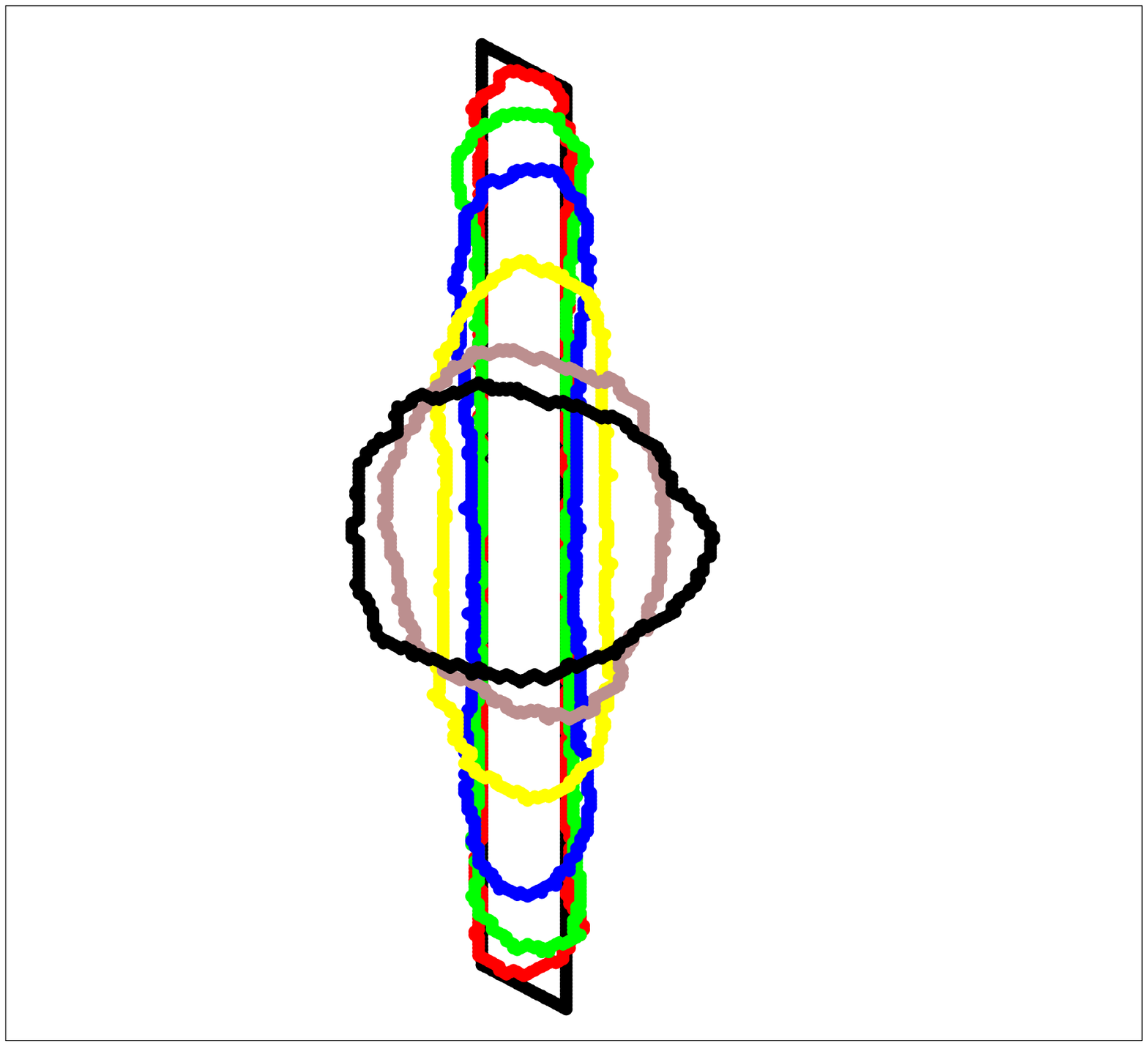}
}}
\end{figure}
\begin{figure}
\centerline{
\hbox{(b)
\epsfxsize=5cm
\epsfbox{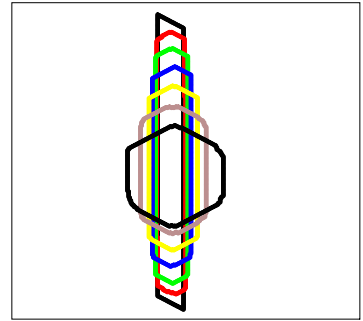}}
}
\caption{Time evolution of islands containing 6250 atoms at 500 K (a) and
83 K (b). The initial state corresponds to the most elongated configuration.}
\label{perimt} 
\end{figure}

\begin{figure}
\centerline{
\hbox{
\epsfxsize=5cm
\epsfbox{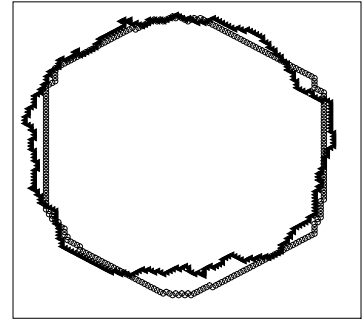}}
}
\caption{Large facets are present at 83 K (open circles) in contrast with
the rugosity observed at 500 K (filled triangles)}
\label{6250fac}
\end{figure}

\begin{figure}
\centerline{
\hbox{(a)
\epsfxsize=5cm
\epsfbox{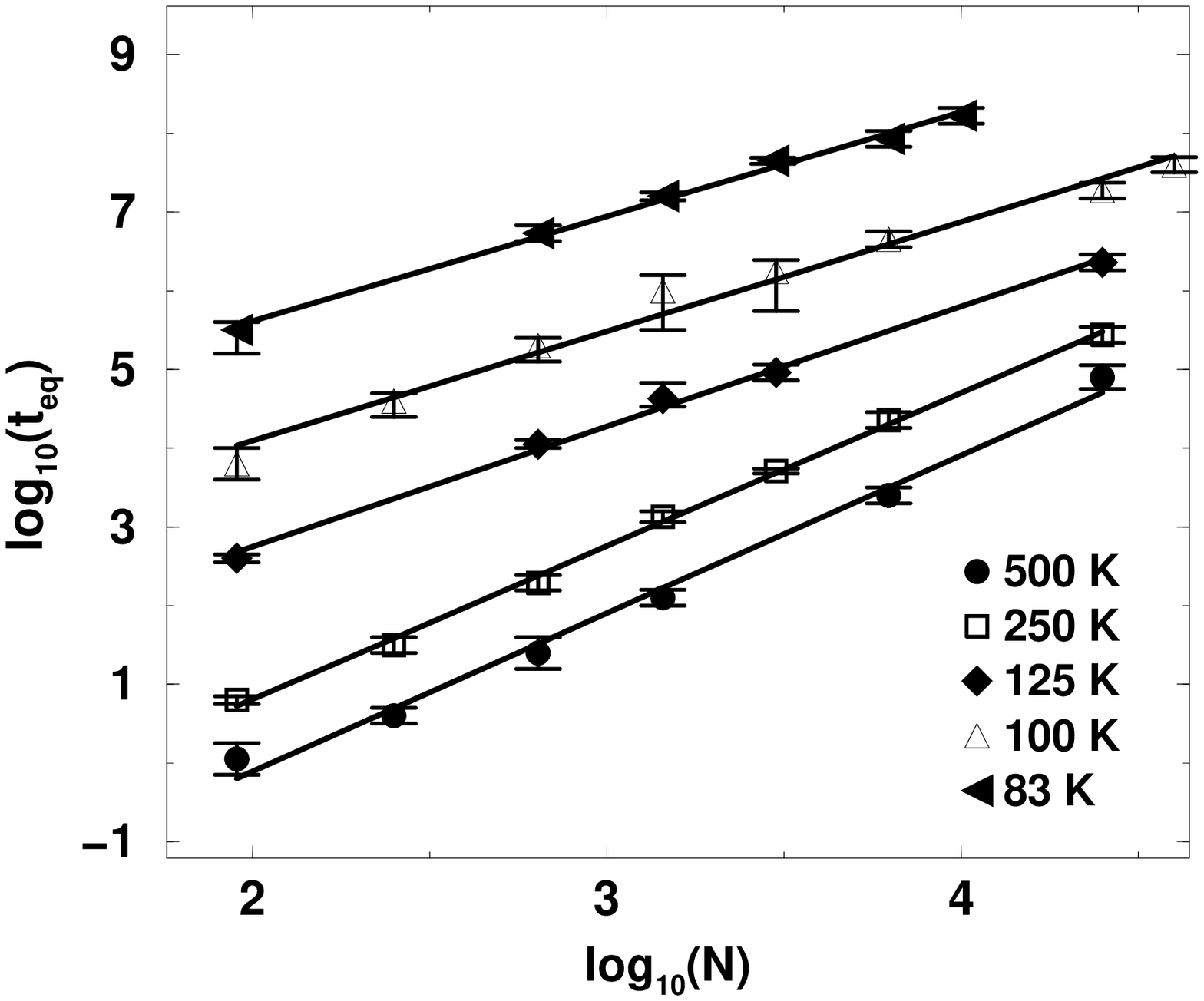}
}}
\end{figure}

\begin{figure}
\centerline{
\hbox{(b)
\epsfxsize=5cm
\epsfbox{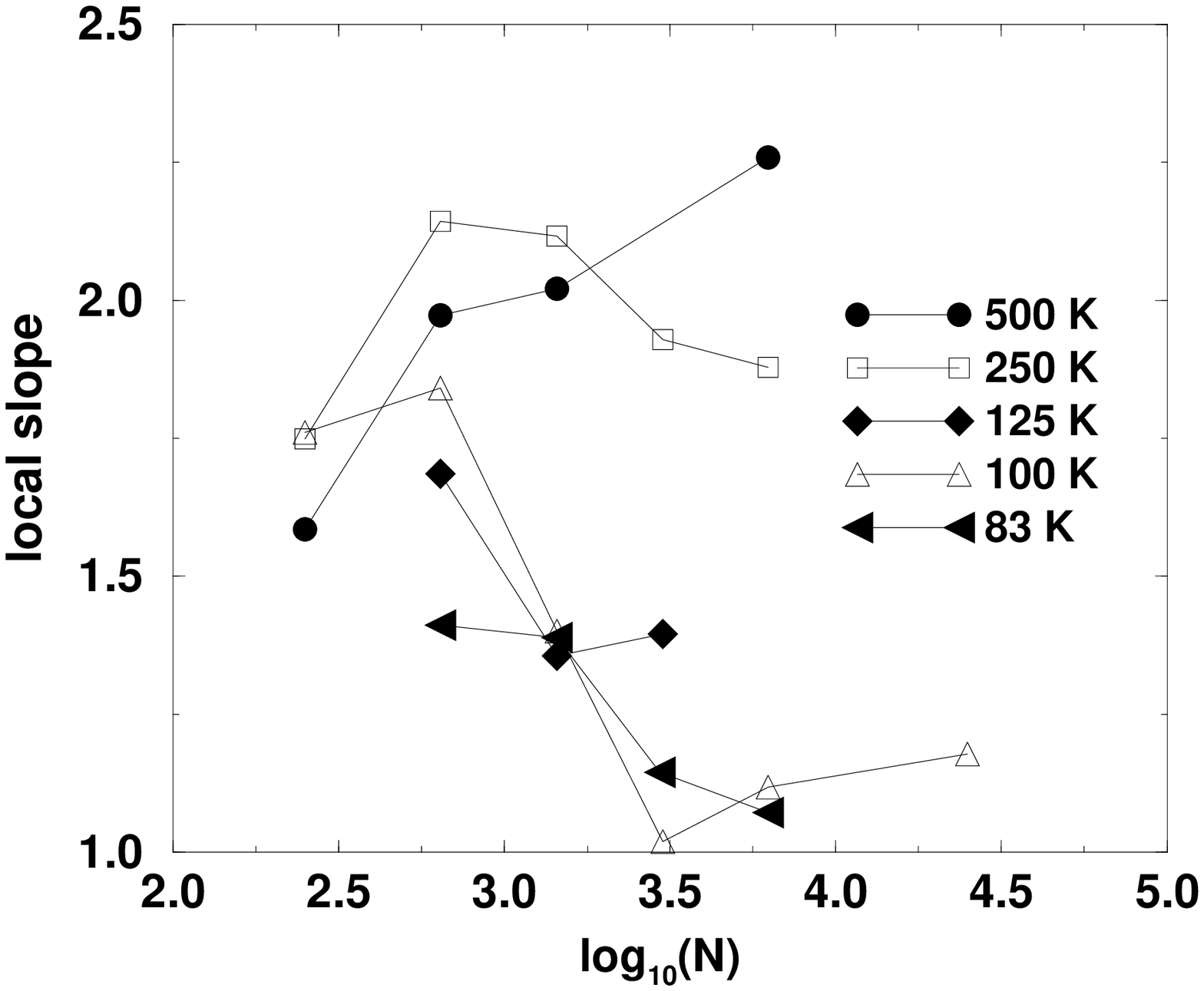}}
} 
\caption{(a) Size dependence of $t_{eq}$ for different temperatures and 
(b) local slope of the size dependence (obtained as the discrete derivative 
of the $t_{eq}$ vs $N$ curve given in (a) : for point i, it is 
$(log(t_{eq}(i+1)) - log(t_{eq}(i-1)))/(log(N(i+1)) - log(N(i-1)))$. 
The local slope remains 
close to 2 at high temperatures for all sizes, but it approaches 1 for 
high sizes at low temperatures. In (a), the
curves have been shifted vertically for clarity. The precise fits are
the following : T=500 K : $t_{eq} = 8 \ 10^{-12} N^{2.00}$; T=250 K :
$t_{eq} = 8 \ 10^{-9} N^{1.95}$; T=125 K : $t_{eq} = 0.51 N^{1.52}$;
T=100 K : $t_{eq} = 6 \ 10^{3} N^{1.39}$; T=83 K : $t_{eq} = 9 \
10^{7} N^{1.32}$}
\label{teqfN}
\end{figure}

\begin{figure}
\centerline{
\hbox{(a)
\epsfxsize=5cm
\epsfbox{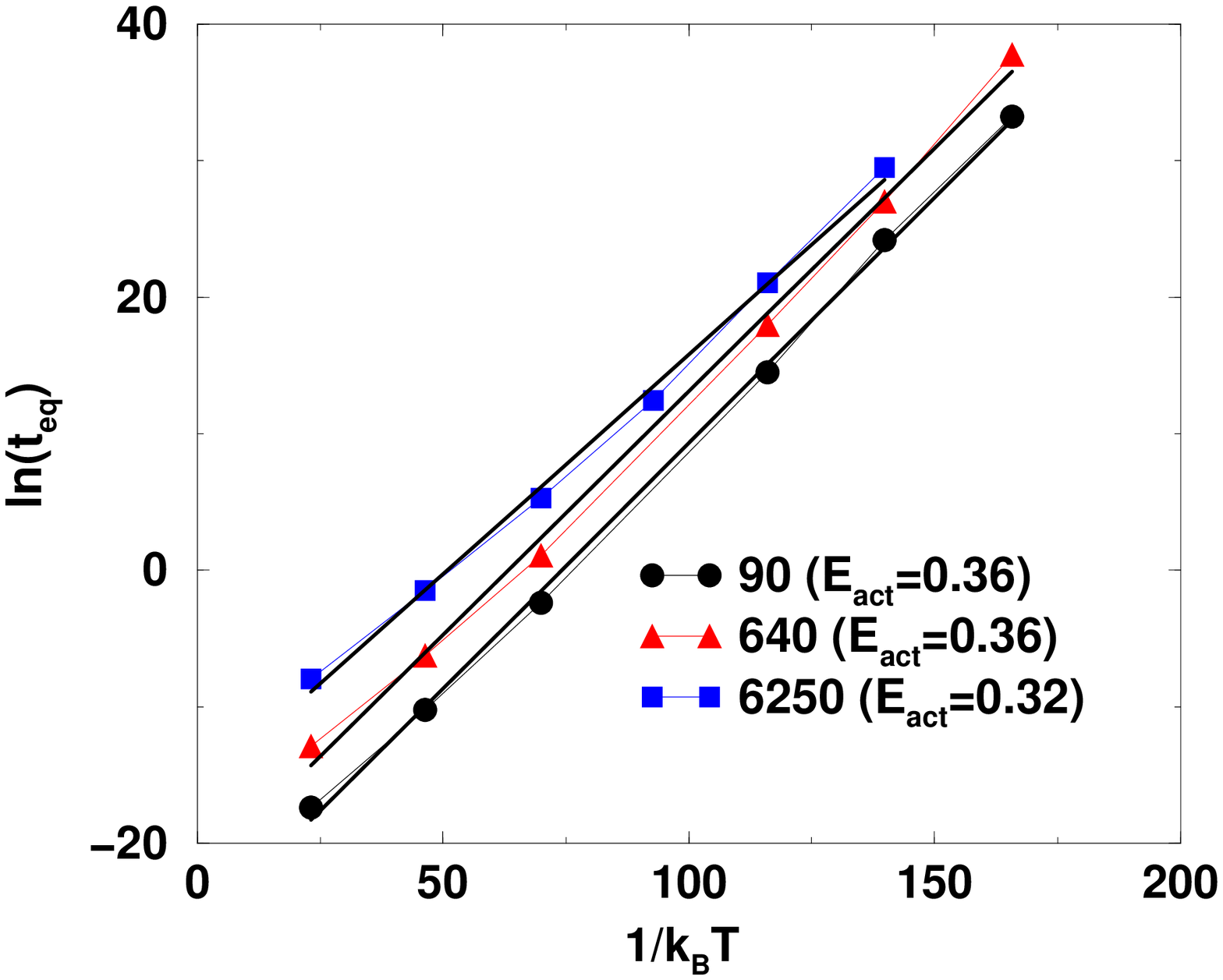}
}}
\end{figure}

\begin{figure}
\centerline{
\hbox{(b)
\epsfxsize=5cm
\epsfbox{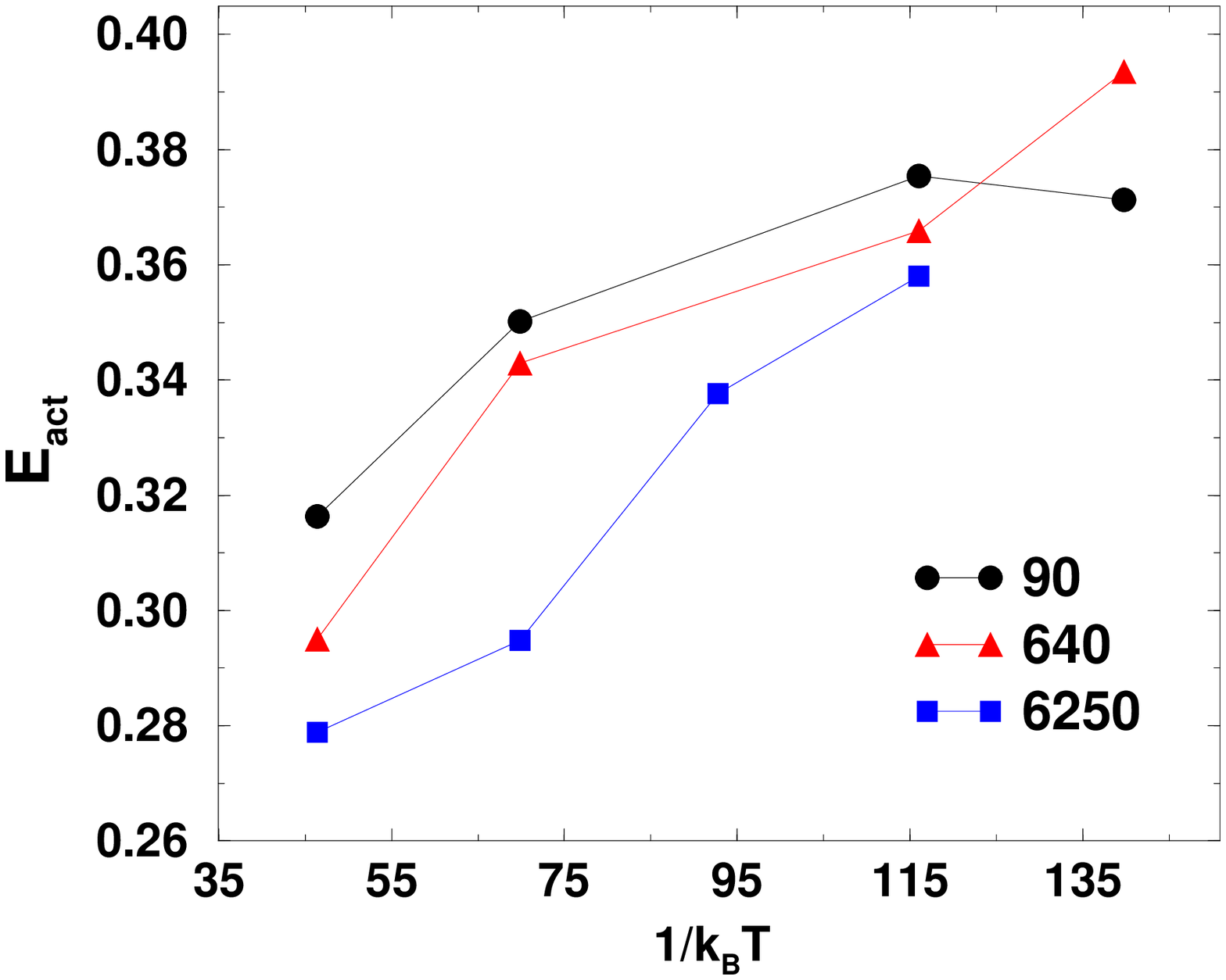}}
} 
\caption{(a) Time needed to reach equilibrium as a function of the
temperature for different island sizes. (b) shows the local
activation energy (defined as in Fig. \protect\ref{teqfN}) at each temperature.}
\label{teqT} 
\end{figure}

\begin{figure}
\centerline{
\hbox{
\epsfxsize=5cm
\epsfbox{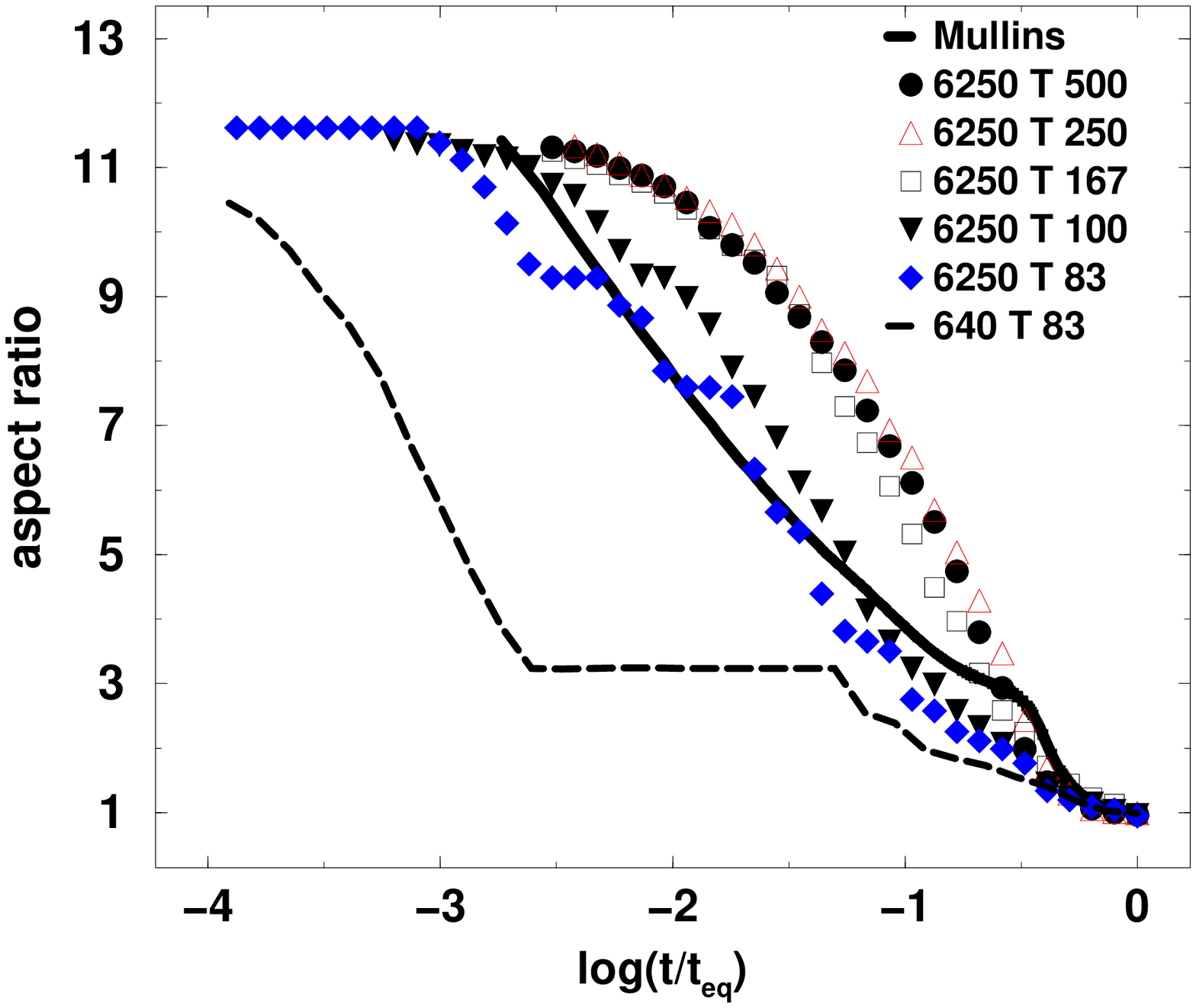}}
}
\caption{Kinetic path followed by the islands as they relax to 
equilibrium.
The solid line represents the prediction of Mullins' equation
\protect\ref{EM2} and the different symbols to the relaxations
obtained in the KMC simulations for islands containing 6250 atoms at
the temperatures shown. The dashed line corresponds to a smaller
island and illustrates the wide range of relaxation paths observed.}
\label{kinetics}
\end{figure}


\begin{thebibliography}{99}

\bibitem{generale} 
M. Lagally, Physics Today {\bf 46}(11), 24 (1993) and references therein;
H. Gleiter, Nanostructured Materials {\bf 1} 1 (1992); 
Z. Zhang and M. G. Lagally, {\it Science} {\bf 276}, 377 (1997)

\bibitem{www_nano}
http://www.msel.nist.gov/structure/metallurgy/techactv95/nanostruc.html
http://nanoweb.mit.edu/
http://www.eas.asu.edu/~nano/nano2.html

\bibitem{fluelli} 
M. Fl\"uelli, P. A. Buffat, and J. P. Borel, Surf. Sci.  {\bf 202},
343 (1988).

\bibitem{duxbury} 
X. Yu and P.M. Duxbury, Phys. Rev B {\bf 52}, 2102 (1995).

\bibitem{metiu} 
H. Shao, S. Liu and H.Metiu, Phys Rev B {\bf 51} 7827 (1995)

\bibitem{averback} 
H. Zhu and R.S. Averback, Phil. Mag. Lett. {\bf 73}, 27 (1996).

\bibitem{LECD}
A. Perez et al, J. of Physics D {\bf 30}, 1 (1997).

\bibitem{bogicevic}
A. Bogicevic, J. Str\"omquist and B. Lundqvist, Phys. Rev. Lett. {\bf 81}, 
637 (1998)

\bibitem{nichols} 
C. Herring, {\it Physics of Powder Metallurgy} (McGraw-Hill Book New 
York, 
Company, Inc. 1951) Ed. W. E. Kingston; Phys Rev {\bf 82} 87 (1951);
C. Herring, Phys Rev {\bf 82} 87 (1951);
W.W. Mullins, J. Appl. Phys. {\bf 28}, 333 (1957) and {\bf 30}, 77 
(1959); 
F.A. Nichols and W.W. Mullins, J. Appl. Phys., {\bf 36}, 1826 (1965); 
F.A. Nichols, J. Appl. Phys. {\bf 37}, 2805 (1966).

\bibitem{drechsler}
M. Drechsler et al. Journal de Physique {\bf 50}, Colloque {\bf C8},
223 (1989)

\bibitem{bonzel}
H. P. Bonzel and E. E. Latta, Surf. Sci. {\bf 76}, 275 (1978)

\bibitem{jeffers}
G. Jeffers, M. A. Dubson and P. M. Duxbury, J. Appl. Phys. 
{\bf 75}, 5016 (1994)

\bibitem{jullien}
N. Olivi-Tran, R. Thouy and R. Jullien, J. Phys I France {\bf 6} 557 
(1996); R. Thouy, N. Olivi-Tran and R. Jullien, Phys. Rev B {\bf 56}, 
5321 (1997)

\bibitem{eggers}
J. Eggers, Phys Rev Lett. {\bf 80} 2634 (1998)

\bibitem{vipi} 
J. Villain and A. Pimpinelli {\it Physique de la Croissance
Cristalline} (Eyrolles, 1995); A. Pimpinelli and J. Villain {\it Physics 
of Crystal Growth} (Cambridge University Press, 1998)

\bibitem{villa}
F. Lançon and J. Villain, in {\it Kinetics of Ordering and 
Growth at Surfaces} M.G. Lagally ed. (Plenum, New York) p. 369 (1990) 

\bibitem{llewis}
L. Lewis, J. L. Barrat and P. Jensen, Phys Rev B {\bf 56} 2248 (1997)

\bibitem{es}
R. L. Schwoebel, J. Appl. Phys. {\bf 40}, 614 (1969); R. L.  Schwoebel
and E. J. Shipsey, J. Appl. Phys. {\bf 37}, 3682 (1966); J. Villain,
J. Physique I {\bf 1}, 19 (1991)

\bibitem{kalos} 
A. B. Bortz, M. H. Kalos and J. L. Lebowitz, J. Comp. Phys.  {\bf 17}
10 (1975)

\bibitem{kang} 
H.C. Kang, W.H.Weinberg, J.Chem.Phys.  {\bf 90}(5), 2824 (1989).

\bibitem{accemetiu} 
Y.T.Lu, H.Metiu, Surf. Science, {\bf 245} 150 (1991).

\bibitem{stoldt} 
C. R. Stoldt, A. M. Cadilhe, C. J. Jenks, J. M. Wen, J. W. Evans and P. A. Thiel, Phys Rev Lett. {\bf 81} 2950 (1998)

\bibitem{miva}
C. Misbah and A. Valance, unpublished

\bibitem{bcf}
W. K. Burton,  N. Cabrera, and F. Frank, {\it Phil. Trans. Roy. Soc.}
{\bf 243}, 299 (1951)  

\bibitem{bz}
G. S. Bales and A. Zangwill, Phys. Rev. B {\bf 41}, 5500 (1990). 

\bibitem{pimpicine}
A. Pimpinelli, J. Villain, D. E. Wolf, J. J. M\'etois, J. C.
Heyraud, I. Elkinani and G. Uimin, Surf. Sci. {\bf 295}, 143 (1993);
A. Pimpinelli, I. Elkinani, A. Karma, C. Misbah and J. 
Villain, J. Phys.: Condens. Matter {\bf 6}, 2661 (1994)

\bibitem{bales}
G.S. Bales and D.C. Chrzan, Phys. Rev. B {\bf 50}, 6057 (1994)

\bibitem{prbevap}
P. Jensen, H. Larralde and A. Pimpinelli, Phys. Rev. B {\bf 55}, 
2556 (1997)

\bibitem{metiumrs} 
H. Metiu (private communication) : more details to be found in 
Metiu, Mattsson and Mills, MRS Symposia (1998), to be published

\bibitem{metiudiff} 
A. Bogicevic, S. Liu, J. Jacobsen, B. Lundqvist and H.Metiu, Phys Rev
B {\bf 57} R9459 (1998)


\end{thebibliography}
\end{document}